\documentclass[submission, Phys]{SciPost}
\usepackage{amsmath,amssymb,mathtools,xspace}
\usepackage{booktabs,multirow,graphicx,tabularx,slashed}
\usepackage{hyperref}
\usepackage{bm}
\usepackage{color,xcolor}
\usepackage[normalem]{ulem}
\usepackage{enumitem}
\usepackage{braket}
\usepackage{stackrel}
\usepackage{tikz}
\usetikzlibrary{arrows}
\usetikzlibrary{shapes.geometric}
\usepackage{float}
\usepackage{subcaption}

\usepackage{epstopdf}
\graphicspath{{figs/}}

\makeatletter
\def\BState{\State\hskip-\ALG@thistlm}
\makeatother

\makeatletter
\@ifundefined{pdfoutput}{}{\DeclareGraphicsRule{*}{mps}{*}{}}
\makeatother

\makeatletter
\DeclareRobustCommand*{\bfseries}{%
   \not@math@alphabet\bfseries\mathbf
   \fontseries\bfdefault\selectfont
   \boldmath
}
\makeatother

\setitemize{itemsep=2pt,topsep=2pt,parsep=0pt,partopsep=0pt,leftmargin=*}
\setenumerate{itemsep=0pt,topsep=2pt,parsep=0pt,partopsep=0pt,labelindent=3pt,leftmargin=*}
\definecolor{Gcolor}{HTML}{3b528b}
\definecolor{Dcolor}{HTML}{e41a1c}

\tikzstyle{generator} = [rectangle, rounded corners, minimum width=3cm, minimum height=1cm,text centered, draw=Gcolor]
\tikzstyle{discriminator} = [rectangle, rounded corners, minimum width=3cm, minimum height=1cm,text centered, draw=Dcolor]
\tikzstyle{io} = [circle, trapezium left angle=70, trapezium right angle=110, minimum width=1cm, minimum height=1cm, text centered, draw=black]

\tikzstyle{process} = [rectangle, minimum width=1cm, minimum height=1cm, text centered, draw=black]
\tikzstyle{decision} = [rectangle, minimum width=1cm, minimum height=1cm, text centered, draw=black]

\tikzstyle{arrow} = [thick,->,>=stealth]
\usepackage{xcolor}

\newcommand\one{\leavevmode\hbox{\small1\normalsize\kern-.33em1}}

\newcommand{\qqquad}{\qquad \qquad}

\def\slashchar#1{\setbox0=\hbox{$#1$}           
   \dimen0=\wd0                                 
   \setbox1=\hbox{/} \dimen1=\wd1               
   \ifdim\dimen0>\dimen1                        
      \rlap{\hbox to \dimen0{\hfil/\hfil}}      
      #1                                        
   \else                                        
      \rlap{\hbox to \dimen1{\hfil$#1$\hfil}}   
      /                                         
   \fi}

\newcommand{\ie}{\textsl{i.e.}\;}

\def\mpl{{m_{\rm{Pl}}}}

\newcommand{\ltsima}{$\; \buildrel < \over \sim \;$}
\newcommand{\lsim}{\lower.5ex\hbox{\ltsima}}
\newcommand{\gtsima}{$\; \buildrel > \over \sim \;$}
\newcommand{\gsim}{\lower.5ex\hbox{\gtsima}}

\newcommand{\dd}{\mathrm{d}}

\begin{document}

\begin{center}
{\Large \textbf{Cornering Extended Starobinsky Inflation with CMB and SKA}}\end{center}

\begin{center}
Tanmoy Modak\textsuperscript{1}, 
Lennart R\"over\textsuperscript{1}, 
Bj\"orn Malte Sch\"afer\textsuperscript{2}, \\
Benedikt Schosser\textsuperscript{1}, and
Tilman Plehn\textsuperscript{1}
\end{center}

\begin{center}
{\bf 1} Institut f\"ur Theoretische Physik, Universit\"at Heidelberg, Germany\\
{\bf 2} Astronomisches Recheninstitut, Zentrum f{\"u}r Astronomie der Universit\"at Heidelberg, Germany\\
\end{center}

\begin{center}
\today
\end{center}

\section*{Abstract}
{\bf Starobinsky inflation is an attractive, fundamental model to
  explain the Planck measurements, and its higher-order extension may
  allow us to probe quantum gravity effects.  We show that future CMB
  data combined with the 21cm intensity map from SKA will meaningfully
  probe such an extended Starobinsky model. A combined analysis will
  provide a precise measurement and intriguing insight into
  inflationary dynamics, even accounting for correlations with
  astrophysical parameters.}


\tikzstyle{int}=[thick,draw, minimum size=2em]


\vspace{10pt}
\noindent\rule{\textwidth}{1pt}
\tableofcontents\thispagestyle{fancy}
\noindent\rule{\textwidth}{1pt}
\vspace{10pt}

\newpage
\section{Introduction}
\label{sec:intro}

Inflation~\cite{Starobinsky:1980te,Sato:1980yn,Guth:1980zm} 
provides a simple and elegant solution to the observed flatness and
horizon problems and naturally explains the absence of exotic
relics. It also seeds primordial density fluctuations, from which the
cosmic large-scale structure evolves. These structures can be 
observed in the cosmic microwave background (CMB)
anisotropies~\cite{Planck:2018vyg,Planck:2018jri} and in the
large-scale distribution of galaxies.

Among inflationary models, Starobinsky or
$R^2$-inflation~\cite{Starobinsky:1980te, Starobinsky:1983zz,
  Vilenkin:1985md,Mijic:1986iv, Maeda:1987xf} is one of the
best-fitting models to data~\cite{Planck:2018jri, Planck:2015sxf,
  Planck:2013jfk} of the early Universe. It simply extends the action
of general relativity (GR) by a quadratic term in the Ricci-scalar.
For the near-scale invariant power spectrum, deviations from GR
manifest themselves primarily in a weak running of the spectral
index. The value of the scalar amplitude and the spectral index
reported by Planck~\cite{Planck:2018jri, Planck:2015sxf,
  Planck:2013jfk} can be accounted for by adjusting the coefficient of
the $R^2$-term.  The extended Starobinsky model with higher-order
curvature modifications is motivated by quantum gravity, but also from
a purely phenomenological point of view~\cite{Saidov:2010wx,
  Huang:2013hsb, Motohashi:2014tra, Asaka:2015vza,Bamba:2015uma,
  Miranda:2017juz, Cheong:2020rao, Rodrigues-da-Silva:2021jab,
  Ivanov:2021chn, Koshelev:2022olc,Shtanov:2022pdx} and it may shed
light on the UV-completion of Einstein gravity. In this paper we
extend the Starobinsky model by an $R^3$-term and study the
constraining power of future cosmological data.

Planck's observations of the cosmic microwave background (CMB)
temperature and polarisation anisotropies have advanced our
understanding of inflation tremendously~\cite{Planck:2018jri}. The
next generation of CMB experiments will further develop this legacy.
We focus on two future CMB experiments,
LiteBIRD~\cite{Matsumura:2013aja, Hazumi:2019lys, LiteBIRD:2022cnt}
and CMB-S4~\cite{CMB-S4:2016ple, Abazajian:2019eic,
  CMB-S4:2020lpa,CMB-S4:2022ght}. The LiteBIRD satellite mission will
detect primordial $B$-mode polarisation with moderate resolution, but
excellent sensitivity. CMB-S4 stands for the next generation of
ground-based detectors, which are going to be installed over the next
decade, with excellent sensitivity and resolution, but limited sky
coverage~\cite{CMB-S4:2016ple, Abazajian:2019eic, CMB-S4:2020lpa,
  CMB-S4:2022ght}.

We supplement the CMB measurements with the 21cm intensity mapping by
the Square Kilometre Array (SKA)~\cite{Cosmology-SWG:2015tjb,
  Furlanetto:2005ax, Furlanetto:2006jb, Loeb:2008dp,
  Mellema:2012ht,Morales:2009gs, Natarajan:2014rra, Pritchard:2011xb,
  Pritchard:2015fia, Bull:2018lat, SKA:2018ckk, Weltman:2018zrl,
  Zaroubi:2012in}, as a second window to primordial structures. We are
primarily interested in the redshift range $z = 8~...~10$ and $k =
0.01~...~0.2$~Mpc$^{-1}$~\cite{Modak:2021zgb}. The combined datasets
well pick up variations in the spectral index to probe the extended
Starobinsky model over a large range of scales. Structure formation at
these scales is described well by linear physics with Gaussian
statistics~\cite{McQuinn:2005hk, Oyama:2015gma, Feix:2019lpo,
  Xu:2020uws}. The low astrophysical systematics due to $X$-ray,
UV-sources \cite{Watkinson:2015vla, Pacucci:2014wwa, Pritchard:2006sq,
  Warszawski:2008pz, Ma:2018ltb} or baryonic feedback
processes~\cite{Kim:2012jx, Geil:2009ee, Barkana:2004zy} allow us to
extract inflationary parameters from 21cm tomography. While we will
use some simplifying assumptions, the modelling of the reionisation
process at high redshift has reached a high degree of
sophistication~\cite{Gnedin:2006uz, Miralda-Escude:1998adl,
  Furlanetto:2004ha, Iliev:2015aia, Trac:2009bt, Shapiro:2008zf} and
takes care of astrophysical processes, which are likewise modelled in
machine learning approaches~\cite{Villanueva-Domingo:2020wpt,
  Hassan:2019cal}.

In Sec.~\ref{sec:forma} we first discuss the details of the
inflationary dynamics, deriving the required equivalent inflationary
potential for extended Starobinsky models using the Einstein-Jordan
duality. We then start with future CMB data and discuss the expected
likelihoods for LiteBIRD and CMB-S4 in Sec.~\ref{sec:lik_cmb} and
results in Sec.~\ref{sec:res_cmb}. In Sec.~\ref{sec:21cmdata} we study
the 21cm intensity mapping by SKA, again detailing the likelihood in
Sec.~\ref{subsec:21cmlik}, followed by a discussion of the modelling
of the neutral hydrogen fraction as a function of redshift as the most
important astrophysical parameter in Sec.~\ref{subsec:xhi}. The
results on probing the extended Starobinsky model with SKA and the
next generation of CMB experiments are discussed in
Sec.~\ref{subsec:res_SKA}. We summarize our results in
Sec.~\ref{sec:summ} and update our results on the slow-roll
parametrization in the Appendix.

\section{Extended Starobinsky model}
\label{sec:forma}

The Starobinsky model~\cite{Starobinsky:1980te,Starobinsky:1983zz} is
one of the simplest inflationary models, yet best-fitting to Planck
data~\cite{Planck:2018jri}. It is defined in the Jordan frame as
\begin{align}
  S_J  =  \frac{1}{2} \int\text{d}^4x \sqrt{-g_J}\: f(R) \; ,
  \label{eq:acJordan}
\end{align}
where $g_J$ denotes the determinant of space-time metric
${g_{\mu\nu}}_J$ with signature convention $(-,+,+,+)$, $M_P= (8\pi
G)^{-1/2}$, and
\begin{align}
  f(R) = M_P^2 \left( R + \frac{1}{6 M^2} R^2 \right) \; ,
\end{align}
with $M^2 >0$. The original Starobinsky model approximates general
$f(R)$ gravity models with an attractor behavior in the large-field
regime, where a single mass parameter $M$ accounts for the observed
nearly-scale invariant power spectrum and spectral
index~\cite{Planck:2018jri}.  Probing an actual inflationary potential
complements results based on an effective reconstruction of
inflationary potentials in the slow-roll
approximation~\cite{Lesgourgues:2007aa,Modak:2021zgb}.  We extend the
original Starobinsky model by a $R^3$-curvature term,
\begin{align}
  f(R) = M_P^2 \left(R + \frac{1}{6 M^2} R^2 + \frac{c}{36 M^4} R^3\right) \; ,
  \label{eq:f(R)}
\end{align}
where $c$ is a dimensionless coefficient, which can be generated by
quantum corrections. Higher-order terms involving derivatives, Ricci
tensors and Riemann tensors typically involve
ghosts~\cite{Stelle:1976gc}, and we neglect them in favor of the
$R^3$-term as a phenomenological window to physics beyond the simple
Starobinsky model.

\begin{figure}[b!]
  \centering
  \includegraphics[width = 0.6\textwidth]{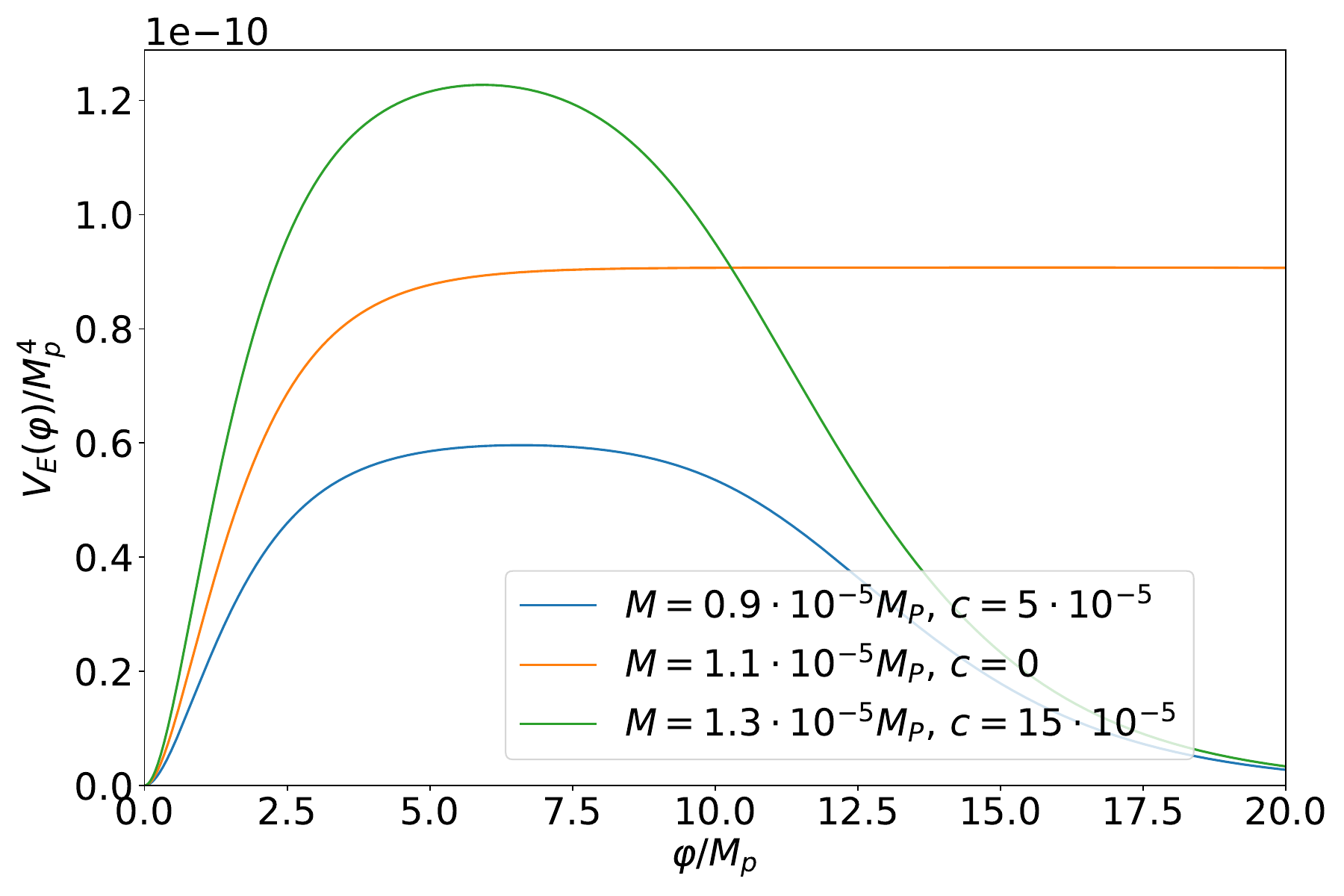}
  \caption{The shape of the inflationary potential for few reference
    choices of $M$ and $c$.}
  \label{fig:pot}
\end{figure}

The corresponding scalar-tensor theory can be found by a Legendre
transformation of Eq.\eqref{eq:acJordan},
\begin{align}
  S_J &= \frac{1}{2} \int\text{d}^4x \sqrt{-g_J}\: \left[ f(s) + f'(s)(R-s)\right]
  \notag \\
  S_J &\equiv   \int\text{d}^4x \sqrt{-g_J}\: \left[\frac{M_P^2}{2} \Omega^2 R - V(s)\right] 
  \notag \\
  & \text{with} \quad \Omega^2 = \frac{f'(s)}{M_P^2}= 1 + \frac{1}{3 M^2} s + \frac{c}{12 M^4} s^2 \notag \\
  & \text{and} \quad \, V(s)=\frac{1}{2}\left[s f'(s)- f(s)\right] \; .
 \label{eq:acJordan2}
\end{align}
The Legendre transform
is well defined as long as $f(R)$ is convex, for Eq.\eqref{eq:f(R)}
translating into $s > -2 M^2 /c$.  The action in
Eq.\eqref{eq:acJordan} can be expressed in the Einstein frame through
the conformal transformation ${g_{\mu\nu}}_E = \Omega^2
{g_{\mu\nu}}_J$,
\begin{align}
 S_E =  \int\text{d}^4x \sqrt{-g_E}\:\left[\frac{M_P^2}{2} R_E -\frac{1}{2} {g^{\mu\nu}}_E
 \left(\nabla_\mu \varphi \nabla_\nu \varphi\right)- V_E(\varphi) \right] \; ,
\end{align}
with the canonical field $\varphi$ and 
\begin{align}
&\varphi = \sqrt{\frac{3}{2}} M_P \ln \Omega^2 ,\label{eq:canofiled}\\
& V_E(\varphi) = \frac{V(s)}{\Omega(s)^4} \Bigg|_{s=s(\varphi)}  \label{eq:canopot},\\
&R = \Omega^2 \left[R_E + 3  \Box_E{\ln\Omega^2}-\frac{3 }{2} g^{\mu\nu}_E \partial_\mu \ln\Omega^2 \ \partial_\nu \ln\Omega^2\right] \; .
  \label{eq:Ricci}
\end{align}
Here, $\Box_E = g^{\mu\nu}_E \partial_\mu \partial_\nu$ is the
d’Alembert operator.  This way, modifications of the gravitational law
are mapped onto an additional field $\varphi$ subjected to dynamics in
a potential $V(\varphi)$. This has the tremendous advantage that the
standard inflationary formalism can be applied for computing the field
dynamics and the associated generation of structures.  In the
potential one has to use $s(\varphi)$, as found by inverting
$\Omega^2$ in Eq.\eqref{eq:canofiled} and solving for
$s(\varphi)$. We find
\begin{align}
    s(\varphi)= 
\begin{cases}
    \dfrac{2 M^2}{c} \left[\sqrt{1+ 3 c (e^{\sqrt{\frac{2}{3}}\frac{\varphi}{M_P}}-1)}-1\right]& \text{for } c\neq 0\\\label{eq:svarphi}
    -3 M^2\left[1- e^{\sqrt{\frac{2}{3}}\frac{\varphi}{M_P}}\right]              & \text{for } c = 0 \; .
\end{cases}
\end{align}
The potential can be expressed as
\begin{align}
&V_E(\varphi) = \frac{M_P^2 \bigg[\dfrac{c s(\varphi)^3}{M^2}+3 s(\varphi)^2\bigg]}{36 M^2 
    \bigg[1+\dfrac{s(\varphi)}{3 M^2}+\dfrac{c s(\varphi)^2}{12 M^4}\bigg]^2} \; .
  \label{eq:potein}
\end{align}
For $c=0$ it can be put into the standard $R^2$ or Starobinsky form 
\begin{align}
V_E (\varphi) = \frac{3 M_P^2 M^2}{4}\left(1- e^{-\sqrt{\frac{2}{3}}\frac{\varphi}{M_P}}\right)^2 \; .
\end{align}
Here $s(\varphi)$ has two solutions, but from Eq.\eqref{eq:svarphi} we
know that we need to satisfy the convexity condition $s > -M^2/(2c)$
and $c >0$, while the potential $V_E(\varphi)$ has to remain positive
at large field values. While the secondary solution can fulfill the
convexity condition for $c < 0$, the potential becomes unbounded from
below for large field values.  In Fig~\ref{fig:pot} we illustrate
$V_E(\varphi)$ for some sample parameter choices.

To study the inflationary dynamics we split $\varphi$ into a classical
background $\bar{\varphi}$ and a perturbation $\delta\varphi$,
\begin{align}
  \varphi(x^\mu) = \bar{\varphi}(t) + \delta\varphi(x^\mu) \; .
  \label{fieldexpan}
\end{align}
The perturbed spatially flat Friedmann-Robertson-Walker (FRW) metric
can be expanded as~\cite{Kodama:1984ziu,Mukhanov:1990me,Malik:2008im}
\begin{align}
\dd s^2 &= -(1+2A) \dd t^2 + 2 a(t) (\partial_i B) \dd x^i \dd t + a(t)^2 \left[(1-2\psi) \delta_{ij}+ 2 h_{ij}\right] \dd x^i \dd x^j,
\end{align}
where $a(t)$ is scale factor and $t$ is the cosmic time. The $A, B,
\psi$ define scalar and $h_{ij}$ tensor metric perturbations.

With the above definitions the background field equation can be
written as
\begin{align}
\ddot{\bar{\varphi}} + 3 H \dot{\bar{\varphi}} + V_{E,\bar{\varphi}} =0,\label{eq:bkg}
\end{align}
where $H= \dd(\ln a)/\dd t$ is the Hubble function fulfilling
\begin{align}
  H^{2} = \frac{1}{3 M_P^2} \left[\frac{1}{2} \dot{\bar{\varphi}}^2 + V_{E}\right]
  \qquad \text{and} \qquad 
  \dot{H}&= -\frac{1}{2 M_P^2} \dot{\bar{\varphi}}^2 \; ,
  \label{hubblesigma2}
\end{align}
The slow-roll parameter $\epsilon$ can then be defined as
\begin{align}
&\epsilon \equiv -\frac{\dot{H}}{H^2} \; .
\end{align}
Inflation ends when $\epsilon = 1$.

Splitting $\varphi(x^\mu)$ into a background field
$\bar{\varphi}(t)$ and gauge-dependent field fluctuations
$\delta\varphi(x^\mu)$ motivates the gauge-independent Mukhanov-Sasaki
variables for the
fluctuations~\cite{Sasaki:1986hm,Mukhanov:1988jd,Mukhanov:1990me,Gong:2011uw},
\begin{align}
  Q = \mathcal{Q} + \frac{\dot{\bar{\varphi}}}{H}\psi
  \qquad \text{with} \qquad 
  \mathcal{Q} = D_\kappa\varphi\vert_{\kappa=0} = \frac{\text{d}\varphi}{\text{d}\kappa}\vert_{\kappa=0} \; ,
\end{align}
where $\kappa$ is the trajectory in field space.  The gauge-invariant
field fluctuations $Q$ fulfill
\begin{align}
&\ddot Q + 3 H \dot Q +\bigg[\frac{k^2}{a^2} + V_{E,\bar{\varphi}\bar{\varphi}}- \frac{1}{M_P^2 a^3} 
    \frac{\text{d}}{\text{d}t}\bigg(\frac{a^3}{H}\dot{\bar{\varphi}}^2\bigg) \bigg] Q=0  \; ,
  \label{eq:fluc}
\end{align}
where $V_{E,\bar{\varphi}\bar{\varphi}}$ is the double derivative of
the potential $V_{E}(\bar{\varphi})$ with respect to $\bar{\varphi}$.
The gauge-invariant curvature perturbation $\mathcal{R}$ is defined
as~\cite{Mukhanov:1990me,Malik:2008im}
\begin{align}
  \mathcal{R} = \frac{ H}{\dot{\bar\varphi}} Q \; ,
  \label{eq:curvpurt}
\end{align}
and we are interested in the power spectrum of the gauge-invariant
curvature perturbation~\cite{Mukhanov:1990me,Bassett:2005xm}
\begin{align}
\langle\mathcal{R}(\bm{k}_1) \mathcal{R}(\bm{k}_2) \rangle = (2\pi)^3 \delta^{(3)}_D(\bm{k}_1+\bm{k}_2) P_{\mathcal{R}}(k_1)
\qquad \text{with} \qquad 
P_{\mathcal{R}}(k)= |\mathcal{R}|^2 \; .
\end{align}
The dimensionless power spectrum for the curvature perturbation is given by
\begin{align}
  \mathcal{P}_{\mathcal{R}}(t;k)= \frac{k^3}{2\pi^2}P_\mathcal{R}(k) \; .
  \label{eq:powadia}
\end{align}
The spectral index $n_{s}$ of the power spectrum of the adiabatic
fluctuations is defined as
\begin{align}
  n_{s} = 1 + \frac{\text{d}\ln\mathcal{P}_{\mathcal{R}}(k)}{\text{d}\ln k} \; .
  \label{specin1}
\end{align}
On the other hand, the mode equation for the tensor amplitude is 
\begin{align}
  v''_{\bm{k}} + \left(k^2 - \frac{a''}{a}\right) v_{\bm{k}} = 0 \; ,
  \label{eq:tensorpert}
\end{align}
where $v_{\bm{k}}$ is the gauge-invariant tensor amplitude and the
prime denotes derivative with respect to conformal time $\tau$ defined
by $dt = a\: d\tau$. The power spectrum of the tensor perturbations is
expressed as
\begin{align}
\mathcal{P}_{\mathcal{T}}(t;k) = 8 \frac{k^3}{2\pi^2} |v_{\bm{k}}|^2 \; .
\end{align}
The tensor-to-scalar ratio $r$, \ie the relative strength between the
tensor and scalar power spectrum evaluated at reference scale $k_* =
0.05$ Mpc$^{-1}$, is defined as
\begin{align}
r = \frac{\mathcal{P}_{\mathcal{T}}}{\mathcal{P}_{\mathcal{R}}}.
\end{align}
To determine the constraints on the Starobinsky model parameters $M$
and $c$ defined in Eq.\eqref{eq:potein} we solve the background and
perturbation equations of Eq.\eqref{eq:bkg}, Eq.\eqref{eq:fluc}, and
Eq.\eqref{eq:tensorpert} in the Cosmic Linear Anisotropy Solving
System (CLASS~III)~\cite{Lesgourgues:2011rg,Blas:2011rf}.

\section{Future CMB data}
\label{sec:cmbdata}

The first data we want to use to probe the inflationary potential are
the CMB anisotropies, which probe the inflationary dynamics through
their sensitivity to the structures in the early Universe. At the relevant
redshifts around $z\simeq 10^3$ the cosmic large scale structure is to
a very good approximation in a state of linear evolution.
Additionally, the relationship between fluctuations in the
gravitational potential, as predicted by linear perturbation theory,
and the observable temperature and polarisation anisotropies is
linear and is not tainted by astrophysics.

\subsection{LiteBIRD and CMB-S4 likelihoods}
\label{sec:lik_cmb}

While we will primarily focus on the future experiments
LiteBIRD~\cite{Matsumura:2013aja,Hazumi:2019lys,LiteBIRD:2022cnt} and
CMB-S4\cite{CMB-S4:2016ple,Abazajian:2019eic,CMB-S4:2020lpa,CMB-S4:2022ght},
we also provide results based on Planck data~\cite{Planck:2018jri} for
validation. Going beyond Planck, future CMB measurements will improve
the probe of small-scale fluctuations, allow better polarisation
measurements, and address the $B$-mode polarisation as an imprint of
tensor fluctuations on large scales. LiteBIRD mainly targets the large
scale for polarisation but lacks sensitivity towards CMB lensing. On
the other hand, CMB-S4 adds on this aspect significantly, except for
large scales, where the small sky fraction and foreground due to lower
sky coverage and fewer channels limits its
reach~\cite{Brinckmann:2018owf}.

We construct Gaussian likelihoods from all four possible spectra,
$C_{TT}(\ell)$, $C_{TE}(\ell)$, $C_{EE}(\ell)$ and $C_{BB}(\ell)$.
They are computed from the input spectra $\mathcal{P}_\mathcal{R}(k)$
and $\mathcal{P}_\mathcal{T}(k)$ which carry information about the
inflationary potential given in Eq.\eqref{eq:potein}, implemented in
CLASS. Each CMB experiment is characterized by its sky fraction, its
instrumental noise, and its angular resolution. They are incorporated
into a covariance, for which we use a Gaussian approximation.

The gravitational lensing effect in the CMB smoothes out the spectra
and, more importantly, converts between $E$-mode and $B$-mode
polarisation.  In our forecasts we assume the lensing effect to be
modelled in the spectra, and we disregard the extracted deflection
angle spectrum $C_{\alpha\alpha}(\ell)$ along with the
cross-correlation $C_{\psi T}(\ell)$ between the lensing potential and
the temperature fluctuation as a source of cosmological
information. In light of the very strong signals from the primordial
fluctuations, gravitational lensing would improve constraints on the
background cosmology and the fluctuation amplitude marginally, but is
not without risk, as the controversy about the Planck lensing
amplitude demonstrated.

The evolution of the scalar and tensor perturbation spectra to the
observable temperature and polarisation spectra of the CMB is handled
by CLASS, and the resulting spectra are assembled into a
$\chi^2$-functional in a Markov Chain Monte Carlo (MCMC) framework
MontePython3~\cite{Brinckmann:2018cvx,Audren:2012wb}. A Markov chain
generates samples from the likelihood
$\mathcal{L}\propto\exp(-\chi^2/2)$ as a function of the fundamental
cosmological parameters, along with the Starobinsky parameters $M$ and
$c$.  While we solve the mode equations for the Starobinsky model, we
consider the subsequent evolution to be governed by standard general
relativity.  The mapping of the Starobinsky model from the Jordan to
the Einstein frame makes the computations of the scalar and tensor
spectra analogous to single-field inflation with a similar
phenomenology of running spectral indices, so we can check our
implementation against the standard $\alpha,\beta$-parametrization for
$\mathcal{P}_\mathcal{R}(k)$.

We use the standard CMB-S4 and LiteBIRD likelihoods in MontePython,
which are described in detail in Ref.~\cite{Brinckmann:2018owf}. For
LiteBIRD the angular scales are $\ell = 2~...~1350$, the sky fraction
is $f_\text{sky}=0.7$, while the channel is taken as 140~GHz with
full-width-half-max or FWHM =~31~arcmin, $\Delta T =4.1$
$\mu$K~arcmin, and $\Delta P =5.8$ $\mu$K~arcmin. The CMB-S4
specifications are $\ell = 30~...~3000$, $f_\text{sky}=0.4$, 150~GHz
channel, FWHM =~3~arcmin, $\Delta T =1.0$ $\mu$K~arcmin and $\Delta P
=1.41$ $\mu$K~arcmin. We need to ensure that the two experiments cover
mutually exclusive $\ell$ ranges, so just as in
Ref.~\cite{Brinckmann:2018owf} we combine low-$\ell$ from LiteBIRD
data and high-$\ell$ CMB-S4 data, separated at $\ell \le 50$. Noise is
estimated through minimum variance estimator for both experiments. We
use the HALOFIT~\cite{Takahashi:2012em} model for the nonlinear
corrections throughout this paper.

\subsection{Combined CMB projections}
\label{sec:res_cmb}

\begin{figure}[t]
  \includegraphics[width = 1\textwidth]{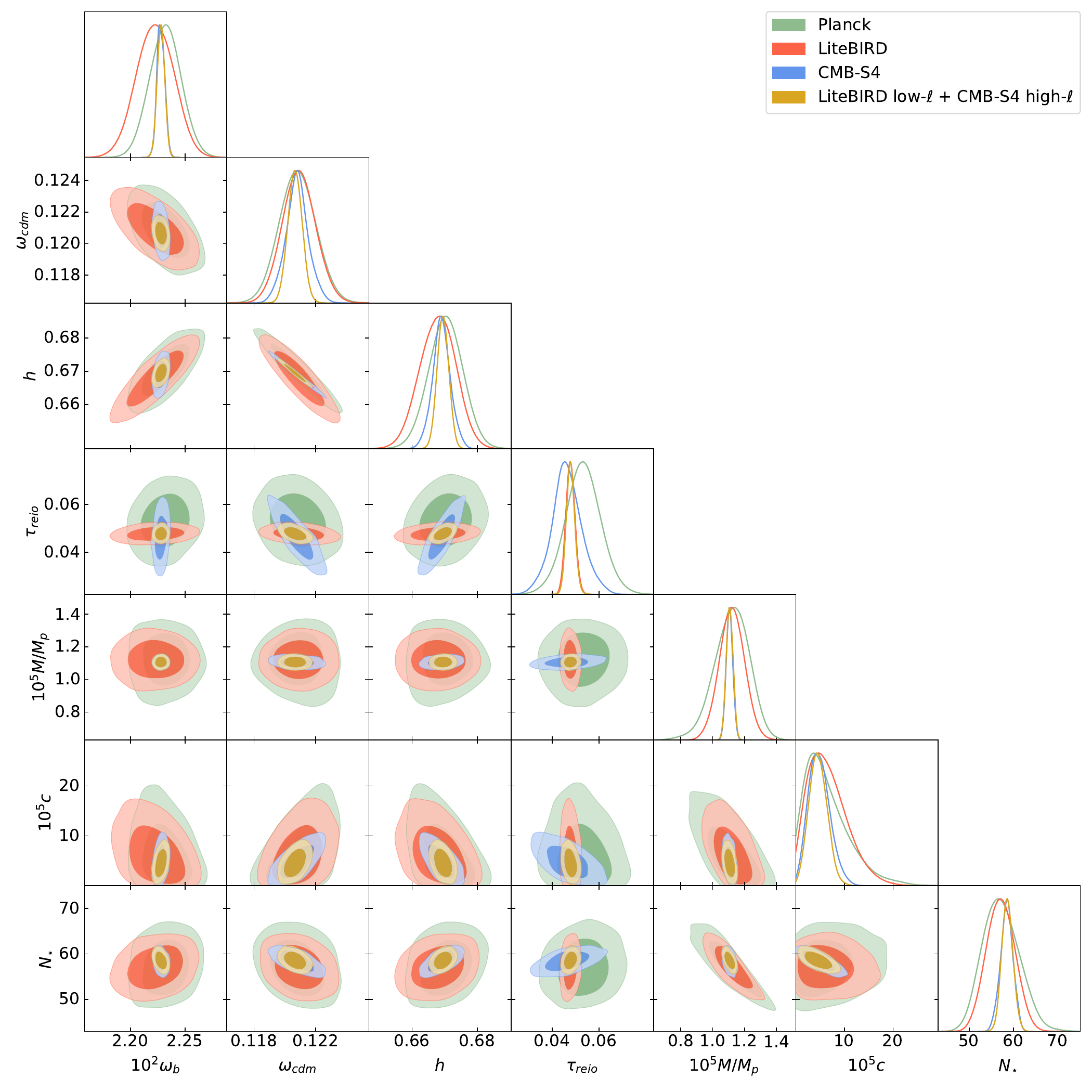}
  \caption{Marginalized CMB posteriors for the extended Starobinsky
    model, based on Planck ($TT$, $TE$, $EE$+low-$\ell$$EE$+low-$\ell$$TT$),
    LiteBIRD, CMB-S4, and the consistent combination of LiteBIRD and
    CMB-S4.}
  \label{fig:extendedstaro_1}
\end{figure}

We use the combined estimated measurements from LiteBIRD and CMB-S4 to
the fundamental parameters $M$ and $c$ in the extended Starobinsky
potential. As the reference cosmological model we choose spatially
flat $\Lambda$CDM-cosmology with parameter space spanned by
\{$\omega_\text{b}$, $\omega_\text{cdm}$, $h$, $\tau_\text{reio}$\},
along with the extended Starobinsky model parameters $\{ M, c\}$, and
$N_*$ as the number of $e$-foldings before the end of inflation, when
the pivot scale $k_*=0.05$~Mpc$^{-1}$ exits the horizon.  We first
consider Planck data, to see what the combined
$TT$, $TE$, $EE$+low-$\ell$$EE$+low-$\ell$$TT$ spectra can tell about $M$ and $c$,
with the baseline model parameters
\begin{align}
\{\; \omega_\text{b} , \omega_\text{cdm} , h , \tau_\text{reio}, M , c , N_* \; \} \; .
\label{eq:model_paras}
\end{align}

For our MCMC runs in MontePython we use the Metropolis-Hastings
algorithm and sample from a Gaussian proposal function with eight
chains totaling up to 5.5 millions steps.  We use flat priors for all
parameters except for $N_*$, for which a Gaussian prior with mean
$\mu_{N_*}=55$ and standard deviation $\sigma_{N_*}= 5$ leads to a
faster convergence of the chains. To check for convergence we use the
criterion $R-1\lesssim 0.05$.  The marginalized posterior
distributions are shown by the green contours in
Fig.~\ref{fig:extendedstaro_1} with the best fit, mean with errors and
corresponding 95\%CL limits given in
Tab.~\ref{tab:Starobinskyext_cmb}.  It is clear from the
Tab.~\ref{tab:Starobinskyext_cmb} that $c$ is compatible with zero,
but showing a mild positive bias.  Our marginalized values are
completely compatible with
Refs.~\cite{Cheong:2020rao,Ivanov:2021chn,Koshelev:2022olc}.

Next, we take the best-fit values from the from Planck data shown
in Fig.~\ref{fig:extendedstaro_1}, specifically including
\begin{align}
  \frac{M}{M_P} = 1.103 \cdot 10^{-5}
  \qquad \text{and} \qquad
  c = 4.135 \cdot 10^{-5} \; ,
  \label{eq:central_points}
\end{align}
and create likelihoods for LiteBIRD and CMB-S4, also discussed in the Appendix.
Even though LiteBIRD and CMB-S4 are both CMB-experiments, their
different focus on angular scales and polarisation renders them
sensitive to cosmological parameters in different ways, as we see in
Fig.~\ref{fig:extendedstaro_1}. The baryon density $\omega_b$ is
extracted from alternating peak heights of the acoustic peaks, so the
large number of multipoles probed by CMB-S4 yields a better
measurement of $\omega_b$. A similar argument applies to the matter
density $\omega_\text{cdm}$, reflected in the sequence of higher order
peaks, where again CMB-S4 has an advantage. For inflation parameters
$M$ and $c$, the much larger $\ell$-values probed by CMB-S4 can also
be seen to make a difference.  In contrast, measuring the optical
depth $\tau_\text{reio}$ requires excellent polarisation sensitivity
on large scales, giving LiteBIRD a clear advantage.  Still, the
results and especially the control over the astrophysics nuisance
parameters of the inflation measurement improves significantly when we
combined LiteBIRD low-$\ell$ with CMB-S4 high-$\ell$ data, allowing us to measure the assumed value $c = 4.135 \cdot 10^{-5}$ to
\begin{align}
  c = (1.015~...~8.3) \cdot 10^{-5}
  \qquad \text{(95\%CL)} \; .
    \label{eq:c_cmb}
\end{align}
We briefly remark that adding lensing data to $TT$, $TE$,
$EE$+low-$\ell$$EE$+low-$\ell$$TT$ only provides minor improvements,
which we do not show. While gravitational lensing of the CMB is
included in our modelling, we do not carry out a lensing
reconstruction, which yields the deflection angle spectra and the
cross-correlation between the lensing potential and the temperature
map~\cite{PhysRevD.67.083002}. Although CMB-lensing is a source of
cosmological information, it is a resource-intensive analysis with
moderate improvements on inflationary constraints. Controlling the
lensing-induced mode conversion between $EE$ and $BB$ is
well-investigated in the literature, and these results also applies to the
Starobinsky case with running spectral indices~\cite{Seljak_2004}.

\begin{table}[t]
  \centering
  \begin{small}
  \begin{tabular}{l|cc|ccc} 
  \toprule
 Data& Parameters & Best-fit & Mean$\pm\sigma$ & 95\% lower & 95\% upper \\ 
 \midrule 
& $100~\omega_b$ &$2.228$ & $2.232_{-0.015}^{+0.015}$ & $2.203$ & $2.26$ \\ 
&$\omega_\text{cdm}$ &$0.1206$ & $0.1208_{-0.0012}^{+0.0012}$ & $0.1185$ & $0.1232$ \\ 
&$h$ &$0.6696$ & $0.6703_{-0.0053}^{+0.0053}$ & $0.6600$ & $0.6808$ \\ 
Planck&$\tau_\text{reio}$ &$0.04781$ & $0.05315_{-0.0077}^{+0.0074}$ & $0.03764$ & $0.0687$ \\ 
\small{($TT$, $TE$,$EE$+low-$\ell$$EE$}&$10^5M/M_P$ &$1.103$ & $1.119^{+0.117}_{- 0.0987}$ & $0.9005$ & $1.329$ \\ 
+low-$\ell$$TT$)&$10^5c$ &$4.135$ & $6.069^{+2.840}_{- 5.402}$ & --- & $< 15.96$ \\ 
&$N_{\star}$ &$58.24$ & $57.17_{- 4.47}^{+3.73}$ & $49.65$ & $65.24$ \\ 
 \midrule 
&$100~\omega_b$ &$2.229$ & $2.223_{-0.017}^{+0.018}$ & $2.190$ & $2.256$ \\ 
&$\omega_\text{cdm}$ &$0.1204$ & $0.1209_{-0.0011}^{+0.001}$ & $0.1188$ & $0.1231$ \\ 
&$h$ &$0.6705$ & $0.6679_{-0.0055}^{+0.0057}$ & $0.657$ & $0.6785$ \\ 
LiteBIRD&$\tau_\text{reio}$ &$0.04735$ & $0.04775_{-0.002}^{+0.002}$ & $0.04391$ & $0.05171$ \\ 
&$10^5M/M_P$ &$1.144$ & $1.121_{- 0.077}^{+0.077}$ & $0.9676$ & $1.273 $ \\ 
&$10^5c$ &$2.633$ & $6.345^{+2.996}_{-4.801}$ & --- & $< 14.62$ \\ 
&$N_*$ &$57.79$ & $57.08_{-3.19}^{+3.18}$ & $51.04$ & $63.27$ \\ 
\midrule
&$100~\omega_b$ &$2.227$ & $2.228_{-0.004}^{+0.004}$ & $2.221$ & $2.235$ \\ 
&$\omega_\text{cdm}$ &$0.121$ & $0.1208_{-0.0007}^{+0.0007}$ & $0.1192$ & $0.1223$ \\ 
&$h$ &$0.6681$ & $0.669_{-0.0027}^{+0.0027}$ & $0.6634$ & $0.6749$ \\ 
CMB-S4&$\tau_\text{reio}$ &$0.04478$ & $0.04634_{-0.0058}^{+0.0064}$ & $0.03258$ & $0.05963$ \\ 
&$10^5M/M_P$ &$1.098$ & $1.105_{-0.021}^{+ 0.021}$ & $1.065$ & $1.145$ \\ 
&$10^5c$ &$5.166$ & $4.794^{+1.923}_{-2.461}$ & $0.7769$ & $9.543$ \\ 
&$N_*$ &$58.44$ & $58.45_{-1.35}^{+ 1.45}$ & $55.66$ & $61.26$ \\
\midrule
&$100~\omega_b$ &$2.227$ & $2.228_{-0.004}^{+0.004}$ & $2.221$ & $2.235$ \\ 
&$\omega_\text{cdm}$ &$0.1206$ & $0.1207_{-0.0005}^{+0.0005}$ & $0.1197$ & $0.1216$ \\ 
LiteBIRD low-$\ell$&$h$ &$0.6696$ & $0.6695_{-0.0018}^{+0.0018}$ & $0.6659$ & $0.673$ \\ 
+&$\tau_\text{reio}$ &$0.04829$ & $0.04779_{-0.0019}^{+0.0017}$ & $0.04425$ & $0.05148$ \\ 
CMB-S4 high-$\ell$&$10^5M/M_P$ &$1.108$ & $1.106_{-0.021}^{+0.022}$ & $1.064$ & $1.147$ \\ 
&$10^5c$ &$4.177$ & $4.573_{-1.944}^{+1.786}$ & $1.015$ & $8.300 $ \\ 
&$N_*$ &$58.71$ & $58.59_{-1.25
}^{+1.24}$ & $56.15$ & $61.08$ \\ 
\bottomrule
  \end{tabular}
  \end{small}
  \caption{Best-fit values, mean, error bars, and 95\%CL limits for
    the parameters shown in Fig.~\ref{fig:extendedstaro_1}.}
\label{tab:Starobinskyext_cmb}
\end{table}

\section{SKA data}
\label{sec:21cmdata}

As a second probe of inflationary dynamics we focus on fluctuations in the
21cm background generated by spin-flip transitions of neutral
hydrogen. The 21cm background is generated at much lower redshifts
around $z \lesssim 10$. This implies that, depending on the redshift
window considered, nonlinearities could become important on small
scales. Intricacies of reionising radiation sources, radiative
transport, and details of the reionising process would then limit our
analysis.  We target $z = 8~...~10$ and employ a simplified model to
compute fluctuations in the 21cm intensity from the statistics of the
matter distribution with weak non-linearities described by the
halo-model.

\subsection{SKA likelihood}
\label{subsec:21cmlik}

As outlined in Sec.~\ref{sec:lik_cmb}, we evolve the spectra of the
scalar and tensor perturbations with CLASS, and in parallel to the
CMB-spectra we compute the density perturbation spectrum $P_\delta(k)$
to model the 21cm-intensity spectrum. The 21cm-spectra depend on the
wave number $k$, the orientation of the modes relative to the line of
sight $\mu$, and the redshift $z$. They are assembled into a
tomographic, redshift-resolved measurement for maximising the sensitivity. The likelihood is a $\chi^2$-functional, constructed
assuming a Gaussian covariance with the experimental characteristics
of SKA. It can be combined with CMB-likelihoods, assuming statistical
independence. Here, a caveat are the integrated Sachs-Wolfe and the
gravitational lensing effects in the CMB, which are generated by
foreground structures that are directly mapped by their 21cm emission, introducing a weak correlation \cite{cross_Tanaka}.

We incorporate details of the 21cm emission through a
redshift-dependent bias parameter as well as a factor taking care of
redshift space distortions induced by peculiar velocities. We model
the reionisation history with a simple 2-parameter model that captures
the global properties of the reionisation process and is verified
against 21cmFAST~\cite{Mesinger:2010ne,Murray:2020trn}.

We follow closely Ref.~\cite{Sprenger:2018tdb} for the evaluation of
21cm power spectrum in our target redshift range.  Assuming a flat-sky
approximation~\cite{Lemos:2017arq,Asgari:2016txw}, the Fourier mode
$\vec{k}$ and the line-of-sight $\vec{r}$ describe the power spectrum
in terms of
\begin{align}
k = \left| \vec{k}\right|
\qquad \text{and} \qquad
\mu = \frac{\vec{k}\cdot\vec{r}}{k r}  \; ,
\end{align}
with the $k$-components $k_\bot = k \sqrt{1 -\mu^2 }$ and
$k_{\parallel} = \mu k$. This gives us
\begin{align}
  P_{21}(k,\mu,z)
  = f_\text{AP}(z) \times f_\text{res}(k,\mu,z) \times f_\text{RSD}(\hat{k},\hat{\mu},z)
  \times b_{21}^2(z) \times P_\delta(\hat{k},z) \; .
 \label{21cmpow}
\end{align}
The wave-number $k$ and the orientation of a mode relative to the line
of sight $\mu$ are derived quantities, as one needs for a given
redshift the angular diameter distance and the Hubble-function which
themselves depend on cosmology. Therefore, it is necessary to
differentiate between the values $k$ and $\mu$ in the cosmological
model probed in our analysis from the assumed-truth or fiducial
parameters describing the assumed cosmology $\hat{k}$ and $\hat{\mu}$.
$P_\delta$ is the matter power spectrum from CDM and baryons and
\begin{align}
  b_{21} &= \overline{\Delta T_b}(z) b_\text{HI}(z) \notag \\
  & \text{with} \quad  
  \overline{\Delta T_b} \simeq 189 \bigg[\frac{H_0 \ (1+z)^2}{H(z)}\bigg]\Omega_\text{HI}(z)\ h \;\text{mK} \; ,
\end{align}
with the mean differential brightness temperature expressed in terms
of the reduced Hubble parameter $h$ defined through $H_0 = h \times
100~\text{km/(s \ MPc)}$. In addition, $b_\text{HI}(z)$ is an, in
principle, redshift-dependent bias. For simplicity we neglect the
redshift dependence in $b_\text{HI}$ and treat it as a nuisance
parameter. The mass density of neutral hydrogen with respect to
critical density is given by
\begin{align}
  \Omega_\text{HI}(z)
  = \frac{\rho_\text{HI}}{\rho_c}
  = \Omega_b (1-Y_P) \left(\frac{H_0}{H(z)}\right)^2 (1+z)^3  \; x_\text{HI}(z) \; ,
\label{eq:omega_hi}
\end{align}
with $\Omega_b = 0.0495$. $Y_P = 0.24672$~\cite{Planck:2018vyg} is the
primordial helium fraction, and $x_\text{HI}(z)$ is the neutral
hydrogen fraction discussed in detail in Sec.~\ref{subsec:xhi}.

Going back to Eq.\eqref{21cmpow}, the so-called Alcock-Paczinsky
effect, or the relative change in the power spectrum between true and
the assumed true (i.e. fiducial) cosmology, is accounted for by
\begin{align}
f_\text{AP}(z) = \frac{D_A^2 \hat{H}}{\hat{D}_A^2 H} \; ,
\end{align}
where $H$ and $D$ are the Hubble parameter and angular diameter
distance as a function of $z$. Quantities within the true cosmology are denoted with $\hat{}$ , e.g. $\hat{H}$. The Fourier-modes are characterised by wave number $k$ and orientation $\mu$ relative to the line of sight, where the relation in these quantities between the true cosmology and and assumed cosmological model is given by
\begin{align}
\hat{k}^2  &= \bigg[\frac{\hat{H}}{H}^2 \mu^2 + \frac{D_A}{\hat{D}_A}(1-\mu^2)\bigg] k^2 \notag \\
\hat{\mu}^2 &= \frac{\hat{H}}{H}^2 \mu^2 \bigg[\frac{\hat{H}}{H}^2 \mu^2 + \frac{D_A}{\hat{D}_A}(1-\mu^2)\bigg]^{-1} \; .
\end{align}
Next, $f_\text{res}(k,\mu,z)$ describes the finite resolution of the
instruments, which suppresses the perturbations on small scales,
\begin{align}
f_\text{res}(k,\mu,z) = 
\exp\left[ -k^2\left(\mu^2 (\sigma_{\parallel}^2- \sigma_{\bot}^2) + \sigma_{\bot}^2\right) \right] \; ,
\end{align}
where $\sigma_{\parallel}$ and $\sigma_{\bot}$ are the Gaussian errors
of the coordinates parallel and perpendicular to the line of sight at
redshift $z$. They are given by
\begin{align}
  \sigma_{\parallel} &= \frac{c}{H}(1+z)^2 \frac{\sigma_\nu}{\nu_0}
  \; &\text{and} \qquad 
  \sigma_{\bot} &= (1+z) D_A \sigma_\theta \notag \\
  \text{with} \qquad 
  \sigma_\theta &= \frac{1}{\sqrt{8 \ln 2}}\frac{\lambda_0}{D_\text{base}}(1+z)
  \; &\text{and} \qquad 
  \sigma_\nu &=\frac{\delta_\nu}{\sqrt{8 \ln 2}} \; .
  \phantom{pushalittle}
\end{align}
The first quantity is the Gaussian suppression of the power spectrum
defined as the ratio between the root mean square and a FWHM of
$\sqrt{8 \ln 2}$. The latter corresponds to the channel width due to
the band separation into different channels with $\lambda_0 =
21.11$~cm, which translates to $\nu_0 = 1420.405752$~MHz.  We use the
SKA1-LOW specifications~\cite{SKA:2018ckk}, expected for observing in
one band $\nu = 50~...~350$~MHz, where the 21cm line in our target
redshift $z = 8~...~10$ lies.  The core SKA1-LOW configuration is an
array of 224 antennas with diameter $D=40$~m and with maximum baseline
$D_\text{base}=1$ km~\cite{SKA:2018ckk}. Here, we use 64000
channels~\cite{skasystem} with $D_\text{base}= 1$~km, again for
SKA1-LOW~\cite{SKA:2018ckk}.
 
Finally, the classical cosmological redshift induces an apparent
anisotropy in the power spectrum, as described by the Kaiser
formula~\cite{Kaiser:1987qv} in the linear regime. Furthermore, the
random peculiar velocities of the galaxies lead to the so-called
fingers-of-God effect~\cite{Jackson:1971sky} in the redshift. Both are
included through the term $f_\text{RSD}$ Eq.\eqref{21cmpow} and described 
by~\cite{Bull:2014rha}
\begin{align}
  f_\text{RSD}(\hat{k},\hat{\mu},z)
  &= \left( 1+ \beta(\hat{k},z) \hat{\mu}^2\right)^2 \;
  e^{-\hat{k}^2 \hat{\mu}^2 \sigma_\text{NL}^2} \notag \\
  \text{with} \qquad &
  \beta(\hat{k},z)
= -\frac{1+z}{2 b_{21}(z)} \frac{\text{d} \log P_\delta(\hat{k},z)}{\text{d} z} \; .
\end{align}
This form of $\beta$ is valid for $k = 0.01~...~0.2~\text{Mpc}^{-1}$
and $z = 8~...~10$.  The first term represents the Kaiser formula, the
exponential term the fingers of God. We take $\sigma_\text{NL} =
1$~Mpc as our fiducial value, which corresponds to non-linear scale of $k_\text{NL}=1$~Mpc$^{-1}$. Due to our conservative $k$-range, this choice has very little effect.
\medskip

The entire observed 21cm power spectrum is a combination of the signal
and noise~\cite{Tegmark:2008au},
\begin{align}
  P^\text{obs}_{21}(k,\mu,z) &=  P_\text{21}(k,\mu,z) + P_N(z) \notag \\
  \text{with} \quad &
   P_N(z)= \frac{4 \pi T_\text{sys}^2 f_\text{sky} \lambda^2 y D_A^2}
   {A \Omega f_\text{cover}  t_\text{obs}} \; .
   \label{nois}
\end{align}
Here $t_\text{obs}$ is the total observation time which we take to be 10000 hrs, $N_\text{dish}$ is
the number of antennas, $f_\text{sky} = 0.58$.  In our analysis we
consider a field of view of $\Omega = (1.2 \lambda/D)^2$, an area $A =
N_\text{dish} \pi (D/2)^2$ per antenna, and the covering fraction
$f_\text{cover}= N_\text{dish}(D/D_\text{base})^2$. Again, we follow
the design specification of SKA1-LOW ~\cite{SKA:2018ckk}.  The system
temperature is the combination of the sky temperature and the receiver
temperature~\cite{SKA:2018ckk}
\begin{align}
  T_\text{sys} = T_\text{sky} &+ T_\text{rx} \notag \\
  \text{with} \quad
  T_\text{sky} &= 25\: \text{K} \left(\frac{408\:\text{MHz}}{\nu}\right)^{2.75} 
  \quad \text{and} \quad 
  T_\text{rx} = 0.1 T_\text{sky} + 40~\text{K} \; ,
\end{align}
and $\nu = \nu_0/(1+z)$. Unlike Ref.~\cite{Sprenger:2018tdb}, where
the noise model treats SKA as a single-dish experiment, our noise
model is based on interferometry. Furthermore, $y$ is defined as
\begin{align}
y =\frac{18.5 \text{MPc}}{1\:\text{MHz}} \left(\frac{1+z}{10}\right)^{1/2} \; .
\end{align}
\medskip

For the 21cm intensity mapping, we divide the mapping into bins of
width $\Delta z$ with mean redshift $\bar{z}$. The volume of one
redshift bin can then be approximated as
\begin{align}
V_r(\bar{z}) = 4 \pi f_\text{sky} \int_{\Delta r(\bar{z})} r^2\text{d}r\: 
= \frac{4 \pi}{3} f_\text{sky} \left[ r^3 \left(\bar{z} +\frac{\Delta z}{2} \right)-r^3 \left(\bar{z} - \frac{\Delta z}{2} \right)\right].
\end{align}
The Gaussian $\chi^2$ giving the likelihood is then defined as the
integral over $k$ and $\mu$ for each redshift band as~\cite{Sprenger:2018tdb}
\begin{align}
\chi^2 = \sum_{\text{bins} \; n} \int_{k_\text{min}}^{k_\text{max}} k^2\text{d}k\: \int_{-1}^{1} \text{d}\mu\: \frac{V_r(\bar{z}_n)}{2(2\pi)^2}
\bigg[\frac{\left(\Delta P_\text{21}(k,\mu,\bar{z}_n)\right)^2}{(P_{21}(k,\mu,\bar{z}_n)+ P_N)^2+
\sigma^2_\text{th}(k,\mu,\bar{z}_n)}\bigg] \; ,
\end{align}
where $\Delta P_{21}$ is the difference between the fiducial and
sampled power spectra, and 
\begin{align}
\sigma_\text{th}(k,\mu,z) = \bigg[\frac{V_r(z)}{2(2\pi)^2} k^2 \Delta\mu \Delta k \frac{\Delta z}{\Delta \bar{z}} \bigg]^{1/2}
\alpha(k,\mu,z) \; P_\text{21}(k,\mu,z) \; .
\label{eq:alpha}
\end{align}  
This uncertainty depends on the correlation lengths $(\Delta k,
\Delta\mu, \Delta z)$.  For a given bin $(k_i, z_j)$, the choice of
$\Delta\mu$ depends on the number of independent nuisance parameters
describing the errors for different $\mu_k$.  Following
Ref.~\cite{Sprenger:2018tdb}, for a given bin $(k_i, z_j)$ the error
on $P_\text{21}(k,\mu,z)$ for different $\mu$ values can be treated as
fully correlated. Taking one parameter per bin is then equivalent to
$\Delta \mu = \mu_\text{max} - \mu_\text{min} \approx 2$ for our
redshift range, reducing Eq.\eqref{eq:alpha} to
\begin{align}
\sigma_\text{th}(k,\mu,z) = \bigg[\frac{V_r(z)}{(2\pi)^2} k^2 \Delta k \frac{\Delta z}{\Delta \bar{z}} \bigg]^{1/2}
\alpha(k,\mu,z) \; P_\text{21}(k,\mu,z) \; .
\label{eq:alpha2}
\end{align}
The correlation length $\Delta k$ is assumed to be $0.05~h\text{/Mpc}$
as a conservative choice, matching the BAO scale.  We also choose
$\Delta z = 1$, which is slightly lower than the whole redshift range
probed by the experiment $z_\text{max}-z_\text{min} = 2$.

The function $\alpha(k,\mu,z)$ accounts for three uncertainties from
different non-linear corrections: The prediction of the matter power
spectrum, the bias, and RSD. Even though non-linear effects are small
in our target redshift range, we include them in our analysis, except
for the RSD source which is negligible for $z = 8~...~10$. The bias is
usually assumed to be linear up to scales $k = 0.2~h\text{/Mpc}$.  The
HALOFIT semi-analytic formula, which we use, includes some of these
effects, but not baryonic and AGN feedback.  To account for the
corresponding uncertainties in the bias and RSD at small scales we
increase the theoretical uncertainties for three reference
points~\cite{Sprenger:2018tdb}, to a 0.33\% error at $k =
0.01~h/\text{Mpc}$, a 1\% error at $k = 0.3~h/\text{Mpc}$, and a 3\%
error at $k = 10~h/\text{Mpc}$.  This translates into
\begin{align}
\alpha(k,z) &=
\begin{cases}
    a_1 \exp\left(c_1 \log_{10} \dfrac{k}{k_1(z)} \right) & \text{for}~ \dfrac{k}{k_1(z)} < 0.3 \\[3mm]
    a_2 \exp\left(c_2 \log_{10} \dfrac{k}{k_1(z)} \right) & \text{for}~ \dfrac{k}{k_1(z)} > 0.3 
\end{cases} \notag \\
k_1(z) &= 1\frac{h}{\text{Mpc}} \left(1+z\right)^{\frac{2}{2+n_s}} \; ,
\label{eq:kcut}
\end{align}
with $a_1 = 1.4806\%$, $a_2 = 2.2047\%$, $c_1 = 0.75056$, and $c_2 =
1.5120$.  As a conservative implementation we apply a sharp cut-off at
$k = 0.2~h/\text{Mpc}$ following the $z$-dependent scaling of
Eq.\eqref{eq:kcut}.

The SKA likelihood is, again, implemented in MontePython, with a
fiducial likelihood based on the best-fit values Planck ($TT$, $TE$,
$EE$+low-$\ell$$EE$+low-$\ell$$TT$) shown in
Tab.~\ref{tab:Starobinskyext_cmb}.  Our updated power spectrum
includes effects which were not considered in our earlier
study~\cite{Modak:2021zgb}, such as the linear biasing factor, the
redshift dependence of neutral hydrogen fraction, $f_\text{AP}(z)$,
and $f_\text{res}$.  The noise model is also significantly improved by
considering the realistic specifications of SKA in the high redshift
region~\cite{SKA:2018ckk}.  As astrophysical inputs for the 21cm power
spectrum we focus on the reionisation history, modelled by the
reionisation redshift, and the velocity at which the Universe
transitions from being neutral to being ionised.  Our modeling is
tested against radiative transfer simulations in Gaussian random
fields with 21cmFAST, confirming that it captures the relevant
physics.

\subsection{Modeling the redshift dependence} 
\label{subsec:xhi}

\begin{figure}[t]
  \includegraphics[width = .5\textwidth]{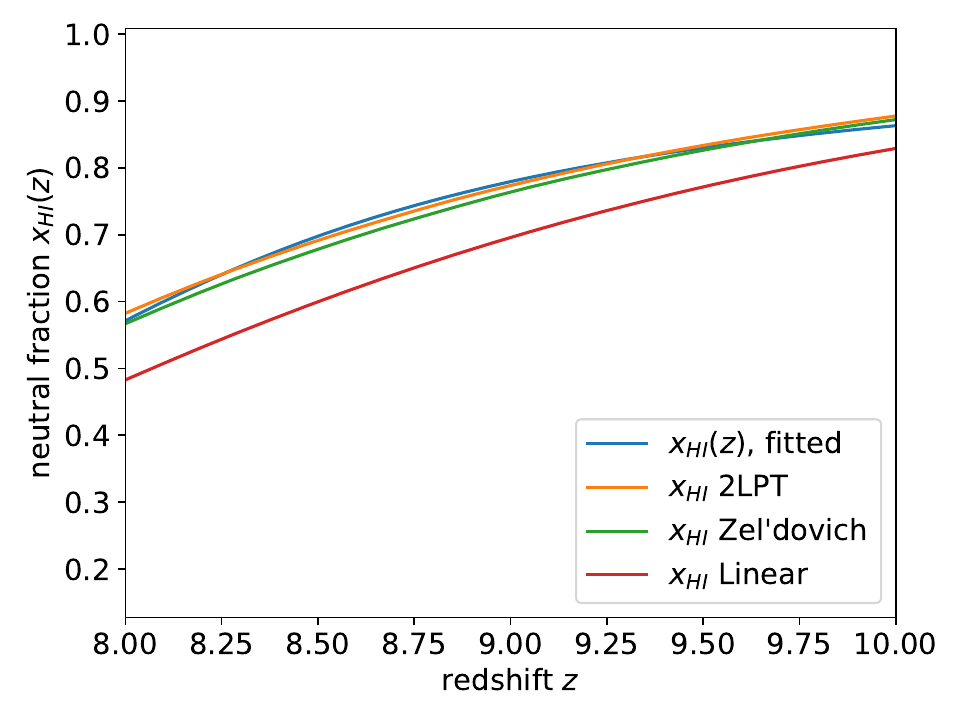}
  \includegraphics[width = .5\textwidth]{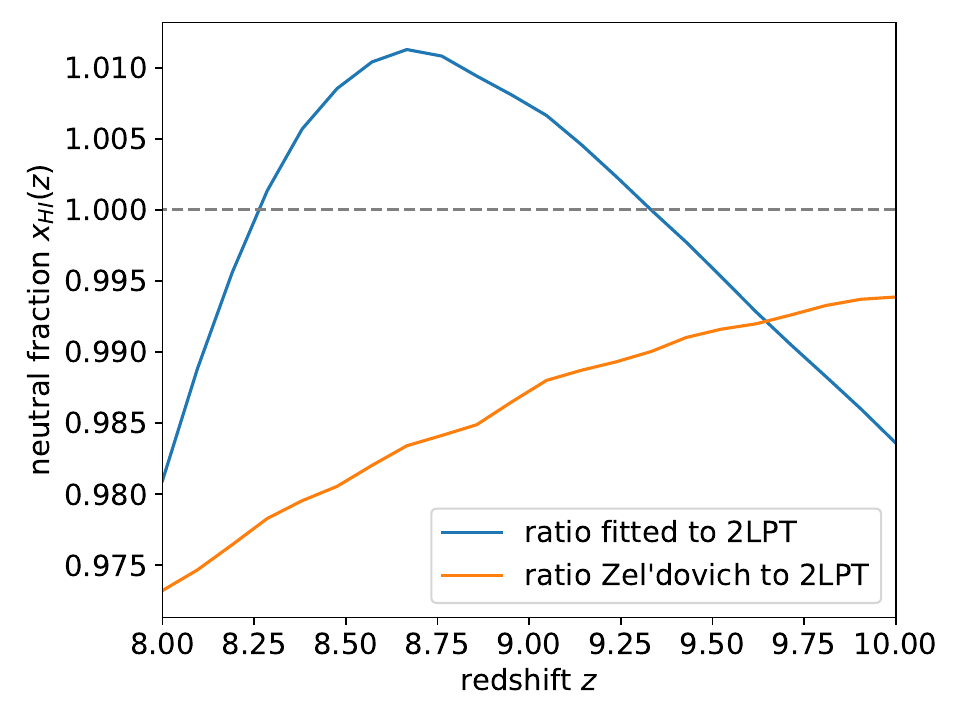}
  \caption{Evolution of the average neutral hydrogen fraction with
    redshift. The fitting function in Eq.\eqref{eq:xHI} yields a
    precision comparable to the Zel'Dovich approximation.}
  \label{fig:neutral_frac}
\end{figure}

To describe the $z$-dependence of $x_\text{HI}$ in
Eq.\eqref{eq:omega_hi} we use the empirical fitting formula
\begin{align}
  x_\text{HI}(z)=\frac{1}{2}\left[ 1+ \frac{2}{\pi} \tan^{-1}\left( \delta_1(z -\delta_2 ) \right) \right] \; ,
  \label{eq:xHI}
\end{align}
where $\delta_1$ and $\delta_2$ are again nuisance parameters.  The
functional shape of Eq.\eqref{eq:xHI} is chosen to fit simulated data
from 21cmFAST~\cite{Mesinger:2010ne,Murray:2020trn}. In our target
redshift region $z = 8~...~10$, the neutral hydrogen fraction is
extracted using the default parameters of 21cmFAST. For each of the 22
linearly spaced redshift bins a cube with side lengths $200$~Mpc is
simulated in real space. The computation is carried out on a
$300\times 300 \times 300$ grid, using the default astrophysics
settings of 21cmFAST. The initial power spectrum is chosen to match
CLASS, which corresponds to the cosmological parameters
$\omega_b=0.02237$, $\omega_\text{cdm} = 0.120$, $h=0.6736$,
$A_s=2.100\cdot 10^{-9}$, $n_s=0.9649$, and $z_\text{reio} =
11.357$. These parameters are computed from the ones defined in Eq.\eqref{eq:model_paras}.

\begin{figure}[b!]
  \includegraphics[width = 1\textwidth]{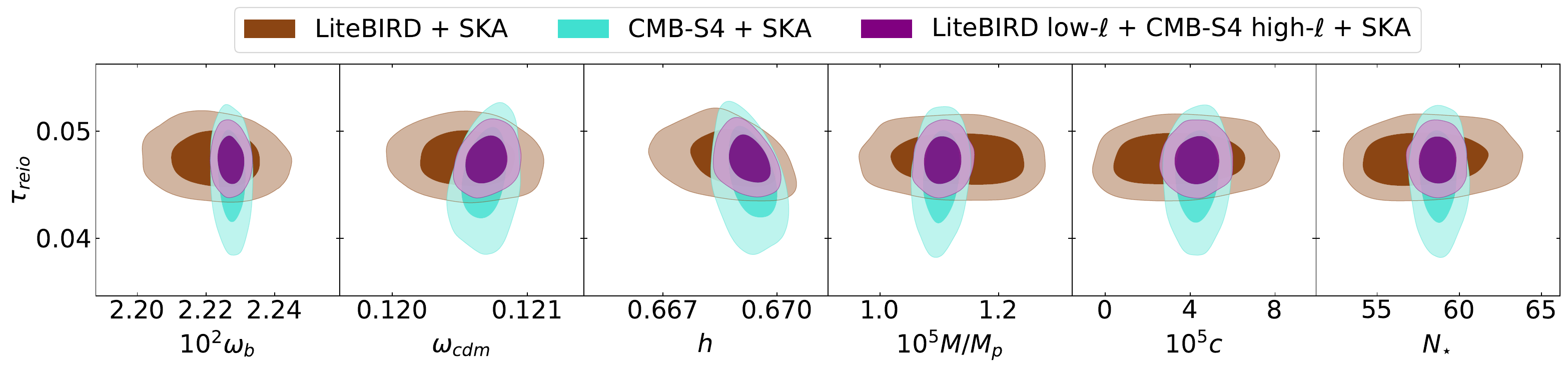}
  \caption{Correlation between $\tau_\text{reio}$ and the remaining
    parameters of Eq.\eqref{eq:model_paras}, based on LiteBIRD,
    CMB-S4, and the combination of LiteBIRD low-$\ell$ and CMB-S4
    high-$\ell$ with SKA.}
  \label{fig:extendedstaro_3}
\end{figure}

\begin{figure}[t]
  \includegraphics[width = 1\textwidth]{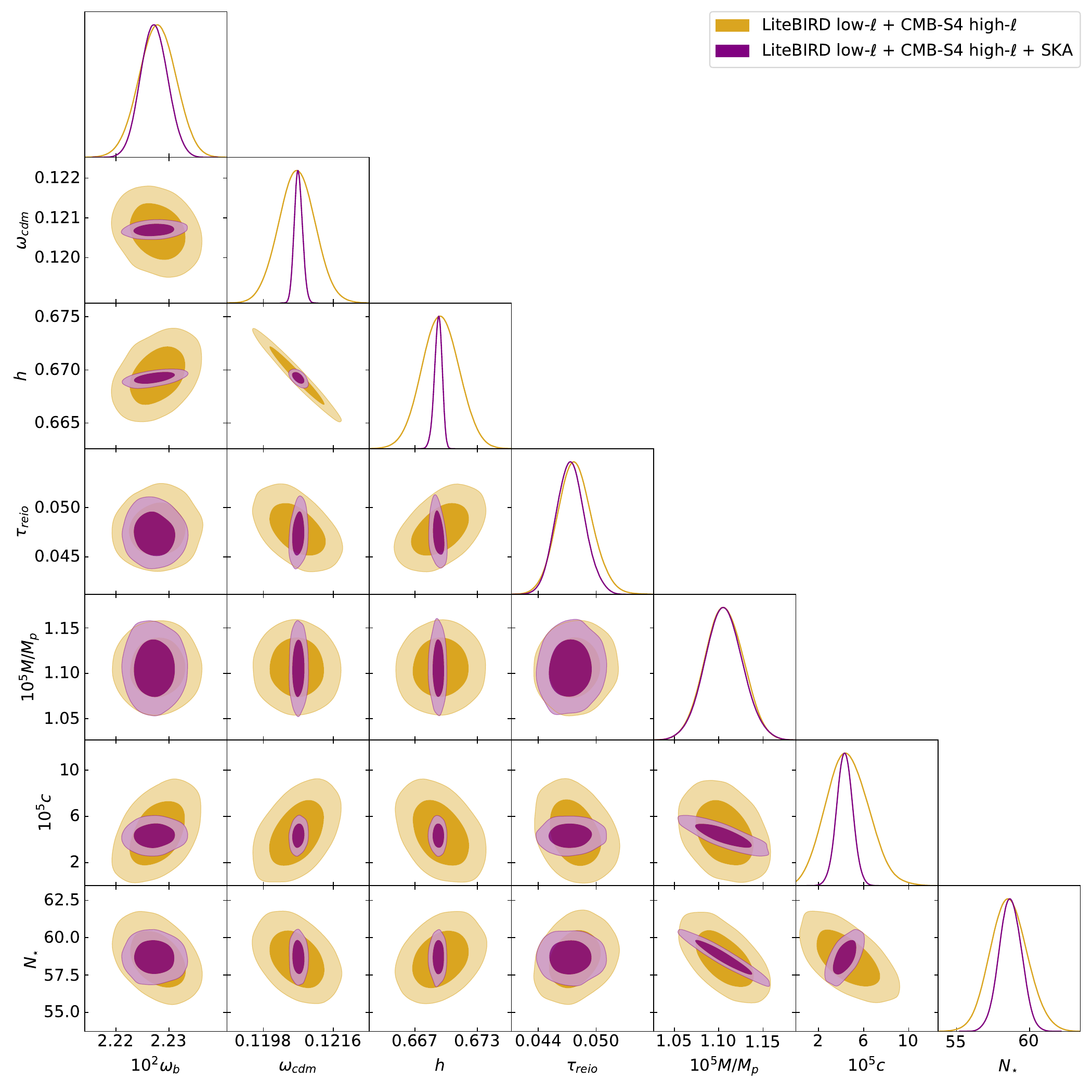}
  \caption{Marginalized CMB and SKA posteriors for the extended
    Starobinsky model, based on the combination of LiteBIRD and CMB-S4
    with high-$\ell$+SKA projections.}
  \label{fig:extendedstaro_2}
\end{figure}

For each of the simulated cubes we compute the average neutral
hydrogen fraction using a first-order perturbative approximation
(Zel'dovich's approximation) and a second-order 2LPT approximation to the linear
velocity field in 21cmFAST.  We then perform a one-dimensional fit of
Eq.\eqref{eq:xHI} to the 2LPT results, giving $\delta_1 = 0.9755$ and
$\delta_2 = 7.7664$ as fiducial values for our MCMC runs.
Fig.~\ref{fig:neutral_frac} illustrates the quality of this
approximation and shows that the relative difference between
Eq.\eqref{eq:xHI} and the 2LPT result is at most $2\%$ in our redshift
region of interest.

\subsection{Combined SKA and CMB projections}
\label{subsec:res_SKA}

We now turn our attention to the sensitivity of SKA to the extended
Starobinsky model parameters.  Without any CMB information the SKA
power spectrum is not sufficient to constrain all parameters given in
Eq.\eqref{eq:model_paras}. However, a combination with the Planck data
is already sufficient to provide a convincing
measurement~\cite{Modak:2021zgb}. Here we ask the more challenging
question, namely what does SKA add to the combination of LiteBIRD and
CMB-S4.  In Fig.~\ref{fig:extendedstaro_3} we compare the combined
sensitivity of SKA with LiteBIRD, CMB-S4 and their low-$\ell$ and
high-$\ell$ combination, respectively. We only show the correlations
of $\tau_\text{reio}$ to the remaining parameters, where we see the
excellent polarisation sensitivity on large scales from LiteBIRD. For
all other parameters there is no additional constraining power from
LiteBIRD and the contours are dominated by CMB-S4.

In Fig.~\ref{fig:extendedstaro_2} we compare the combined sensitivity
of LiteBIRD low-$\ell$, CMB-S4 high-$\ell$, and SKA with the CMB
sensitivity alone.  The corresponding best-fit, mean and corresponding
95\%CL limits are given in Tab.~\ref{tab:Starobinskyext_2}.  The
astrophysical parameters benefiting significantly from SKA are
$\omega_\text{cdm}$ and $h$.  While we are mainly interested in the
fundamental parameters of the inflation potential, this kind of
improvement leads to a big improvement in the global analysis. While
the combination with SKA still leaves a narrow correlation between the
astrophysical $N_*$ and the Starobinsky parameter $M$, it provides an
improved reach in the second Starobinsky parameter, as compared to the
CMB projection of Eq.\eqref{eq:c_cmb},
\begin{align}
  c = (2.89~...~5.73) \cdot 10^{-5}
  \qquad \text{(95\%CL)} \; .
  \label{eq:c_ska}
\end{align}
The narrow correlations between $N_*$ and $M$ and, to some extent, $c$
trace back to how Eq.\eqref{eq:bkg} is solved. The initial conditions
to solve Eq.\eqref{eq:bkg} in CLASS require the number of $e$-foldings
before the end of inflation when the reference mode exited the horizon
\ie $N_*$ and the magnitude of $M$ and $c$. This solution is then used
to match the observables $A_s$ and $n_s$, and leads to the strong
correlation found above. Such a correlation can perhaps be resolved
with the better description of the (p)reheating process after
inflation.

\begin{table}[t]
  \centering
  \begin{small}
  \begin{tabular}{l|cc|ccc} 
  \toprule
  Data& Parameters & Best-fit & Mean$\pm\sigma$ & 95\% lower & 95\% upper \\ 
&&&&&\\
 \midrule 

&$100~\omega_b$ &$2.228$ & $2.227_{-0.003}^{+0.003}$ & $2.222$ & $2.232$ \\ 
LiteBIRD low-$\ell$&$\omega_\text{cdm}$ &$0.1206$ & $0.1207_{-0.0001}^{+0.0001}$ & $0.1205$ & $0.1209$ \\ 
+&$h$ &$0.6694$ & $0.6692_{-0.0003}^{+0.0004}$ & $0.6685$ & $0.670$ \\ 
CMB-S4 high-$\ell$&$\tau_\text{reio}$ &$0.04792$ & $0.04734_{-0.0016}^{+0.0014}$ & $0.04445$ & $0.05033$ \\ 
+&$10^5M/M_P$ &$1.100$ & $1.106_{-0.023}^{+ 0.023}$ & $1.064$ & $1.148 $ \\ 
SKA&$10^5c$ &$4.350$ & $4.325_{-0.690}^{+0.692}$ & $2.891$ & $5.734 $ \\ 
&$N_*$ &$58.95$ & $58.68_{-0.75}^{+0.77}$ & $57.20$ & $60.18$ \\ 
\bottomrule
\end{tabular} 
  \end{small}
  \caption{Best-fit values, mean, error bars, and 95\%CL limits for
    the parameters shown in Fig.~\ref{fig:extendedstaro_2}.}
\label{tab:Starobinskyext_2}
\end{table}

\section{Outlook}
\label{sec:summ}

We have estimated the sensitivity of future CMB and SKA measurements
to the Starobinsky model for inflation, extended by a $R^3$-term. Such
a term may hint at physics beyond general relativity, including
quantum gravity.  Planck data prefers a finite $R^2$-terms and
constrains the coefficient of the $R^3$-term to be $c\lesssim
1.6\times10^{-4}$ at 95\%CL.

We performed a global Markov chain analysis, combining astrophysical
and cosmological parameters with the two fundamental parameters
describing Starobinsky inflation.  First, we found that future CMB
data from LiteBIRD and CMB-S4 will constrain the astrophysical
parameters and also the inflationary parameters $M$ and $c$. In
particular, we found that combining the two experiments in mutually
exclusive $\ell$ ranges can probe the coefficient of $R^3$ at the level $c
=(1.01~...~8.3) \times10^{-5}$ at 95\%CL. The assumed finite central
value is given by the best-fit value from our Planck analysis.

Next, we showed that 21cm intensity mapping by SKA will add to the
constraints from CMB data, focusing on the redshift region $z =
8~...~10$.  While the combination of future CMB and SKA data still
leaves us with a sizeable correlation between the number of
$e$-foldings $N_*$ and the scalaron mass $M$, it improves the
measurement of the extended Starobinsky parameters to $c =
(2.9~...~5.7) \times10^{-5}$.  If $c$ is non-zero, SKA will allow for
a robust determination of this fundamental parameter pointing to
physics beyond standard GR.

\subsection*{Acknowledgments}

TM thanks Sung Mook Lee, Kin-ya Oda, and Tomo Takahashi for fruitful
discussions.  TM is supported by Postdoctoral Research Fellowship from
Alexander von Humboldt Foundation.  The research of TP is supported by
the Deutsche Forschungsgemeinschaft (DFG, German Research Foundation)
under grant 396021762 - TRR 257 Particle Physics Phenomenology after
the Higgs Discovery. This work was supported by the Deutsche
Forschungsgemeinschaft (DFG, German Research Foundation) under
Germany’s Excellence Strategy EXC 2181/1 - 390900948 (the Heidelberg
STRUCTURES Excellence Cluster).

\appendix
\section{HSR projections}
\label{sec:HSR}

Finally, we briefly revisit our previous results on the combination of
Planck and SKA~\cite{Modak:2021zgb} and determine the the sensitivity
of future CMB data in terms of the so-called Hubble slow-roll (HSR)
parameters~\cite{Lesgourgues:2007aa,Planck:2018jri}.  In this
parametrization, the inflationary dynamics are captured by
reconstructing the Hubble function in the observable window, defined
by the range of observationally accessible spatial scales as
\begin{align}
  H(\varphi)
  = \sum_{n = 0}^{N} \frac{1}{n!} \;
  \frac{\dd^n H}{\dd\varphi^n} \Bigg|_{\bar{\varphi_*}} (\bar\varphi-\bar{\varphi_*})^n \;.
  \label{eq:recohubble}
\end{align}
To avoid degeneracies it is convenient to use the logarithmic changes
to the Hubble function through the
parameters~\cite{Lesgourgues:2007aa,Planck:2018jri}
\begin{align}
  \lambda^{(n)}_H = \left(\frac{\mpl^2}{4\pi}\right)^n
  \left(\frac{(H^\prime)^{n-1}}{H^n}\frac{\dd^{n+1}H}{\dd\varphi^{n+1}}\right)
  \qqquad  n\geq 1 \; ,
\end{align}
with the correspondence $\eta_H = \lambda^{(1)}$, $\xi^2_H =
\lambda^{(2)}$, and $\omega^3_H = \lambda^{(3)}$.  As in before, we
assume spatially flat $\Lambda$CDM-cosmology with the baseline model,
as described by \{$\omega_\text{b}$, $\omega_\text{cdm}$, $h$,
$\tau_\text{reio}$, $n_s$, $\tilde{A}_s$, $\epsilon_H$, $\eta_H$,
$\xi^2_H$, $\omega^3_H$\}, where we truncate the HSRs after
$\omega^3_H$. The purpose of this Appendix is to investigate the power
of future CMB data in constraining HSRs, along with a new SKA
likelihood with an improved signal and noise modeling as compared to
Ref.~\cite{Modak:2021zgb}.

\begin{figure}[htbp]
  \includegraphics[width = 1\textwidth]{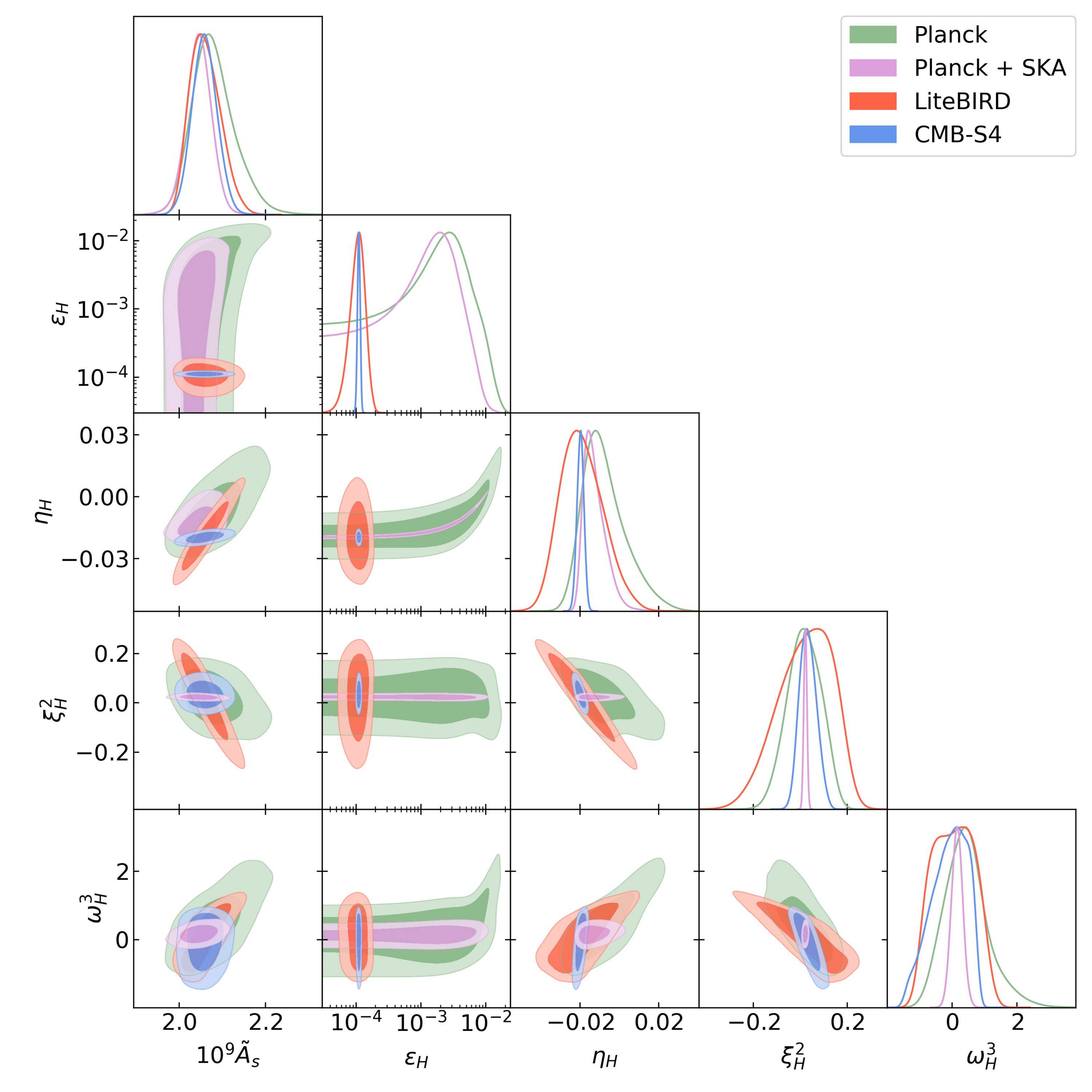}
  \caption{Marginalized CMB posteriors for the HSR parameters based on
    Planck ($TT$, $TE$, $EE$+low-$\ell$$EE$+low-$\ell$$TT$),
    Planck+SKA, LiteBIRD, CMB-S4. }
  \label{fig:HSR_1}
\end{figure}

\begin{figure}[htbp]
  \includegraphics[width = 1\textwidth]{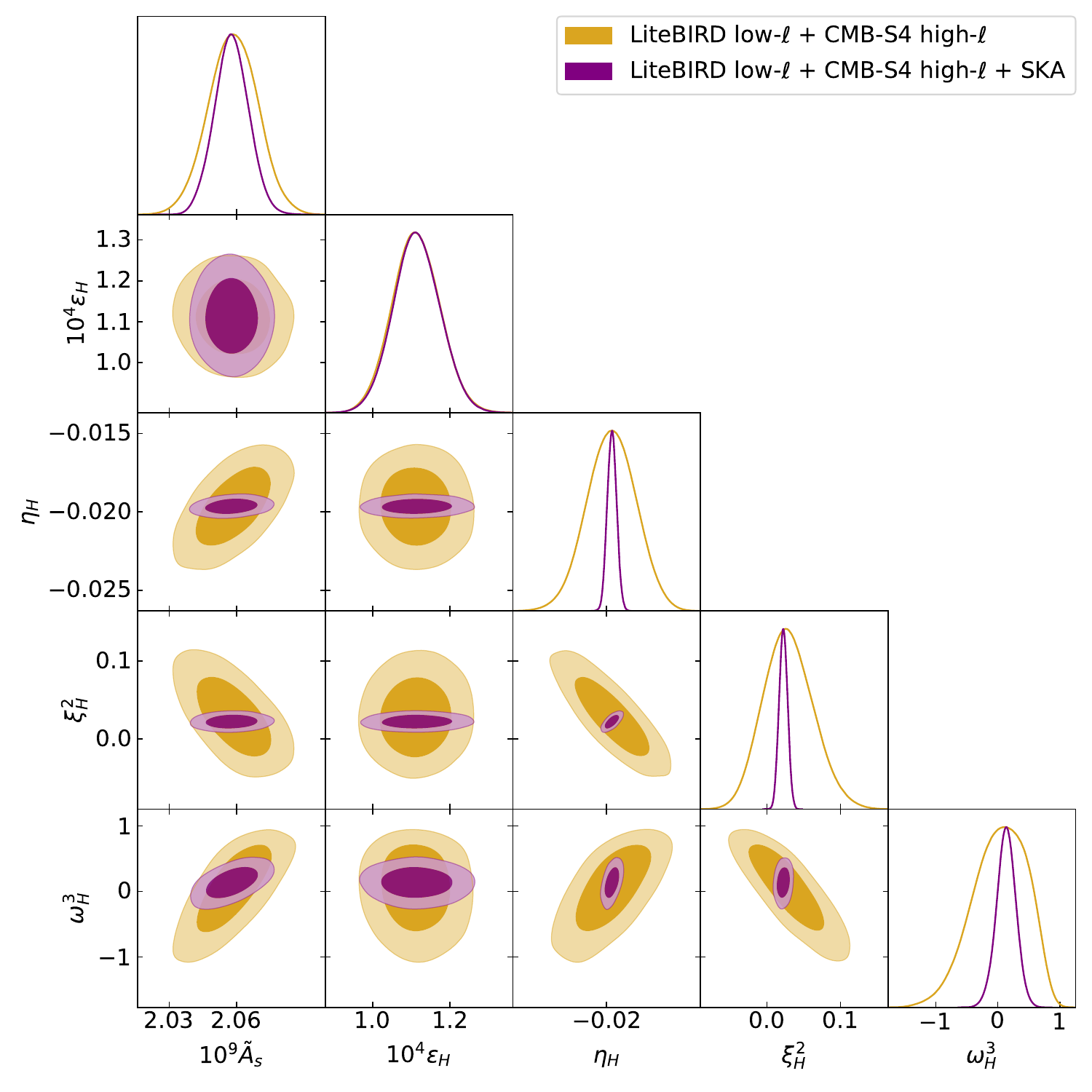}
  \caption{Marginalized CMB posteriors for the HSR parameters based on
    LiteBIRD low-$\ell$+CMB-S4 high-$\ell$ and LiteBIRD
    low-$\ell$+CMB-S4 high-$\ell$+SKA. }
  \label{fig:HSR_2}
\end{figure}

To determine the projected constraints on the HSR parameters we rely
on CLASS and MontePython, as discussed in the main body of the paper.
The expected constraints are shown for Planck, LiteBIRD, CMB-S4, and
Planck+SKA in Fig~\ref{fig:HSR_1} while, for LiteBIRD
low-$\ell$+CMB-S4 high-$\ell$ and LiteBIRD low-$\ell$+CMB-S4
high-$\ell$+SKA they are shown in Fig.\ref{fig:HSR_2}.  The respective
best-fit and mean values are given in Tab.~\ref{tab:HSR}. As in the
extended Starobinsky model, both LiteBIRD+CMB-S4 and
LiteBIRD+CMB-S4+SKA data will provide the best constraints.  We note
that the fiducial likelihoods for the LiteBIRD, CMB-S4 and SKA are
generated with the best-fit values to the marginalized posterior of
Planck $TT$, $TE$, $EE$+low-$\ell$ $EE$+low-$\ell$ $TT$ data, also given
in Tab.~\ref{tab:HSR}.

\begin{table}[htbp]
  \centering
  \begin{small}
   \begin{tabular}{l|cc|ccc} 
     \toprule
 Data& Parameters & Best-fit & Mean$\pm\sigma$ & 95\% lower & 95\% upper \\ 
 \midrule
&$10^{9}\tilde{A}_s$ &$2.059$ & $2.080_{-0.057}^{+0.039}$ & $1.987$ & $2.186$ \\ 
&$\epsilon_H$ &$0.0001111$ & $0.005445_{-0.005313}^{+0.002930}$ & --- & $< 0.01393$ \\ 
Planck&$\eta_H$ &$-0.01953$ & $-0.007755_{-0.012983}^{+ 0.007628}$ & $-0.02712$ & $0.01599$ \\ 
&$\xi^2_H$ &$0.02493$ & $0.01809_{- 0.07272}^{+0.07614}$ & $-0.1175$ & $0.1587$ \\ 
&$\omega^3_H$ &$0.1008$ & $0.4431_{-0.7216}^{+0.5342}$ & $-0.8662$ & $1.812$ \\ 
  \midrule
&$10^{9}\tilde{A}_s$ &$2.034$ & $2.047_{-0.029}^{+0.028}$ & $1.986$ & $ 2.102$ \\ 
Planck&$\epsilon_H$ &$0.001220$ & $0.003338_{-0.003156}^{+0.001678}$ & --- & $< 0.009013$ 
\\ +&$\eta_H$ &$-0.0175$ & $-0.01283_{-0.00658}^{+0.00353}$ & $-0.02158$ & $-0.001015$ \\ 
SKA&$\xi^2_H$ &$0.02386$ & $0.02238_{-0.00676}^{+0.00687}$ & $0.00856$ & $0.0356$ \\ 
&$\omega^3_H$ &$0.01487$ & $0.1594_{-0.1722}^{+0.5060}$ & $-0.1846$ & $0.5023$ \\
\midrule
  &$10^{9}\tilde{A}_s$ &$2.052$ & $2.061_{-0.042}^{+0.030}$ & $1.995$ & $2.131$ \\ 
&$10^4\epsilon_H$ &$1.086$ & $1.151_{-0.316}^{+0.260}$ & $0.5912$ & $1.736$ \\ 
LiteBIRD&$\eta_H$ &$-0.01991$ & $-0.01897_{-0.01229}^{+0.00938}$ & $-0.03931$ & $0.002793$ \\ 
&$\xi^2_H$ &$0.04889$ & $0.02338_{-0.09424}^{+0.13927}$ & $-0.2021$ & $0.2296$ \\ 
&$\omega^3_H$ &$-0.2849$ & $0.04961_{-0.68982}^{+0.67578}$ & $-1.059$ & $1.155$ \\ 
\midrule 
&$10^{9}\tilde{A}_s$ &$2.06$ & $2.059_{-0.030}^{+0.028}$ & $2.002$ & $2.118$ \\ 
&$10^4\epsilon_H$ &$1.104$ & $1.113_{-0.057}^{+0.053}$ & $1.007$ & $1.221$ \\ 
CMB-S4&
$\eta_H$ &$-0.0191$ & $-0.0198_{-0.0018}^{+0.0018}$ & $-0.02334$ & $-0.01632$ \\ 
&$\xi^2_H$ &$0.01098$ & $0.03418_{-0.03942}^{+0.03331}$ & $-0.03461$ & $0.1062$ \\ 
&$\omega^3_H$ &$0.3613$ & $-0.09232_{-0.37747}^{+0.73948}$ & $-1.157$ & $0.8509$ \\ 
\midrule 
&$10^{9}\tilde{A}_s$ &$2.053$ & $2.059_{-0.011}^{+0.011}$ & $2.037$ & $2.08$ \\ 
LiteBIRD low-$\ell$&$10^4 \epsilon_H$ &$1.109$ & $1.112_{-0.064}^{+0.060}$ & $0.9925$ & $1.235$ \\ 
+&$\eta_H$ &$-0.01999$ & $-0.01966_{-0.00165}^{+0.00166}$ & $-0.02291$ & $-0.01636$ \\ 
CMB-S4 high-$\ell$& $\xi^2_H$ &$0.02682$ & $0.0288_{-0.0359}^{+0.0316}$ & $-0.03628$ & $0.09762$ \\ 
&$\omega^3_H$ &$-0.003998$ & $0.03371_{-0.37005}^{+0.51047}$ & $-0.7958$ & $0.8296$ \\ 
\midrule
LiteBIRD low-$\ell$&$10^{9}\tilde{A}_s$ &$2.06$ & $2.058_{-0.008}^{+0.008}$ & $2.042$ & $2.073$ \\ 
+&$10^4 \epsilon_H$ &$1.114$ & $1.114_{-0.062}^{+0.060}$ & $0.9967$ & $1.235$ \\ 
CMB-S4 high-$\ell$&$\eta_H$ &$-0.01951$ & $-0.01964_{-0.00033}^{+0.00031}$ & $-0.02026$ & $-0.01902$ \\ 
+&$\xi^2_H$ &$0.02687$ & $0.02238_{-0.00568}^{+0.00568}$ & $0.01128$ & $0.03339$ \\ 
SKA&$\omega^3_H$ &$0.1544$ & $0.1347_{-0.1525}^{+0.1561}$ & $-0.1974$ & $0.4407$ \\
\bottomrule
 \end{tabular} 
 \end{small}
   \caption{Best-fit values, mean, error bars, and 95\%CL limits for
     the HSR parameters shown in Figs.~\ref{fig:HSR_1}
     and~\ref{fig:HSR_2}.}
 \label{tab:HSR}
 \end{table}

\bibliography{literature}
\end{document}